\documentclass[aps,prb,twocolumn,noeprint]{revtex4-2}

\usepackage{amsmath}
\usepackage{amsfonts}
\usepackage{amssymb}
\usepackage{graphicx}
\usepackage[utf8]{inputenc}
\usepackage[colorlinks=true,urlcolor=blue,linkcolor=blue,citecolor=blue,bookmarks=false]{hyperref}% add hypertext capabilities
\usepackage{color}
\usepackage{braket}
\usepackage{dsfont}
\usepackage{MnSymbol}
\usepackage{xspace}
\usepackage{bbold}
\usepackage{diagbox}
\usepackage{setspace}
\usepackage{soul}

\setcitestyle{square,numbers}

\DeclareMathOperator{\Tr}{Tr}

\newcommand{\ie}[0]{i.e.\@\xspace}
\newcommand{\eg}[0]{e.g.\@\xspace}

\newfont{\tensy}{cmsy10}

\newcommand{\ham}[1]{\hat{H}_{#1}}

\renewcommand{\S}[0]{\hat{\mathcal{H}}}

\newcommand{\ban}[1]{\hat{a}^{\vphantom\dagger}_{#1}}
\newcommand{\bcr}[1]{\hat{a}^{\dagger}_{#1}}

\newcommand{\bop}[2]{\hat{a}^{#2}_{#1}}

\newcommand{\varrhoan}[1]{\hat{\varrho}^{\vphantom\dagger}_{#1}}
\newcommand{\varrhocr}[1]{\hat{\varrho}^{\dagger}_{#1}}

\newcommand{\kB}{k_{\text{B}}}

\newcommand{\im}{\mathrm{i}}

\newcommand{\absolute}[1]{\left| #1 \right|}
\newcommand{\expv}[1]{\left\langle #1 \right\rangle}
\newcommand{\expvtext}[1]{\langle #1 \rangle}

\newcommand{\expvbtext}[1]{\langle #1 \rangle_\text{b}}

\newcommand{\expvb}[1]{\left\langle #1 \right\rangle_\text{b}}

\newcommand{\conf}{\mathcal{C}}
\newcommand{\confn}{{\mathcal{C}_n}}

\newcommand{\spinx}{\hat{S}_{x}}
\newcommand{\spiny}{\hat{S}_{y}}
\newcommand{\spinz}{\hat{S}_{z}}
\newcommand{\spinp}{\hat{S}_{+}}
\newcommand{\spinm}{\hat{S}_{-}}
\newcommand{\spinpm}{\hat{S}_{\pm}}
\newcommand{\spinc}[1]{\hat{S}_{#1}}

\newcommand{\typeint}{t_\mathrm{int}}

\begin{document}

\title{Quantum Monte Carlo simulation of spin-boson models using wormhole updates}

\author{Manuel Weber}
%\email[Email: ]{mweber@pks.mpg.de}
\affiliation{\mbox{Max-Planck-Institut f\"ur Physik komplexer Systeme, N\"othnitzer Str.~38, 01187 Dresden, Germany}}

\date{\today}

\begin{abstract}
We present an exact quantum Monte Carlo method for spin systems coupled to dissipative bosonic baths
which makes use of nonlocal wormhole updates to simulate the retarded spin-flip interactions
originating from an off-diagonal spin-boson coupling. The method is closely related to the stochastic
series expansion and extends the scope of the global directed-loop updates to nonlocal moves
through a world-line configuration. We test our method for the U(1)-symmetric two-bath
spin-boson model, where the off-diagonal components of a spin-$1/2$ particle are coupled to
identical independent baths with power-law spectra, and get a precise estimate of the critical coupling between the critical and the localized phase.
Our method applies to impurity systems and lattice models in any spatial dimension
coupled to bosonic modes with arbitrary spectral distributions. 
\end{abstract}

%\pacs{}
%\keywords{}

%\maketitle must follow title, authors, abstract, \pacs, and \keywords
\maketitle

\section{Introduction}

The development of global updating schemes for Monte Carlo methods
\cite{PhysRevLett.58.86, PhysRevLett.62.361, PhysRevLett.70.875, doi:10.1080/0001873021000049195}
has enabled high-precision numerical studies of critical phenomena
that appear in a wide range of strongly correlated physics.
For quantum systems, the worm and directed-loop algorithms
\cite{Prokofev:1998aa,PhysRevB.59.R14157,PhysRevE.66.046701}
have become the state of the art to simulate either bosonic
or spin models, respectively. While these methods stochastically
sample the perturbation expansion of the partition function,
the worm/directed-loop updates
change the world-line configurations along a closed loop that is constructed
in an extended configuration space. The computational cost to construct 
these loops
scales only linearly in system size and inverse temperature
and therefore allows for efficient 
simulations in any spatial dimension,
as long as the sign problem is absent.
Consequently, it is highly desirable to extend these algorithms
to new classes of models which cannot be simulated efficiently otherwise.

A major challenge in computational many-particle theory is the
simulation of quantum systems coupled to local bosonic modes that do not
obey particle-number conservation.
Among the most studied examples are
lattice models with a local coupling
to phonons. Matrix product state (MPS) based approaches require large
bond dimensions to deal with
the unbound bosonic Hilbert space of each mode, making them less efficient
than for purely electronic or spin models; however,
optimized basis sets can lead
to notable improvements
\cite{STOLPP2021108106}.
 Quantum Monte Carlo (QMC) simulations often suffer from long autocorrelation
 times when the bosons are sampled in first quantization
 \cite{Hohenadler2008}.
 The absence of particle-number conservation in their second-quantized form
 so far inhibited a simple and efficient formulation of the worm/directed-loop updates
 (for exceptions see Refs.~\onlinecite{Assaad2008,SuwaPhD}),
 leaving inefficient local updates as the only available option to simulate
 the bosons \cite{PhysRevB.56.14510, PhysRevB.75.245103}.
 The difficulties of direct boson sampling can be avoided if the Hamiltonian is quadratic
 in the bosonic fields. Then, the bosons can be integrated out exactly using the path integral
 and we obtain a retarded interaction in the system's degrees of freedom.
 A recent generalization of the directed-loop algorithm to retarded interactions \cite{PhysRevLett.119.097401}
 has been shown to overcome the autocorrelation problem and was successfully applied
 to phonon-coupled systems in one \cite{PhysRevB.98.235117, PhysRevResearch.2.023013} and
  two dimensions \cite{PhysRevB.103.L041105}.

 The coupling to bosonic modes plays an important role in many areas of quantum physics, \eg,
 in solid-state systems 
 bosonic excitations appear in the form of quasiparticles like phonons or magnons,
 trapped-ion quantum simulators rely on long-range interactions mediated
 by the vibrational modes of the ions \cite{PhysRevLett.92.207901,RevModPhys.93.025001}, and in quantum optics
 the atoms interact with the quantized electromagnetic field  \cite{scully_zubairy_1997, RevModPhys.80.885}.
 A central problem is the dissipation of a quantum system when
 connected to a bosonic bath with an infinite number of degrees of freedom \cite{RevModPhys.59.1}.
 One of its simplest realizations is the 
 \textit{spin-boson model}: a
 two-level system that is coupled to a continuum of bosonic modes.
  In this model, the competition between bath-induced localization
  and delocalization due to an applied magnetic field leads to a quantum
  phase transition
  whose critical properties
  had only been resolved with the development
  of an efficient cluster QMC method that simulates the coupling
  to the bath in terms of a diagonal
  retarded interaction \cite{PhysRevLett.102.030601}. 
  A variational MPS approach revealed that the coupling
  of two noncommuting spin components to independent baths
  can lead to even more complex phase
  diagrams \cite{PhysRevLett.108.160401}
  including a critical phase \cite{Sachdev2479, PhysRevB.61.4041}
  that emerges due to frustration effects in the decoherence mechanism \cite{PhysRevLett.91.096401};
  however, generalizations of the MPS approach to multiple baths
  are limited by the large dimensions of local bosonic Hilbert spaces.
  Lattice models with onsite dissipation have been simulated using classical
  Monte Carlo methods for long-range interactions
  \cite{1995IJMPC...6..359L, 2005PThPS.160..395W};
  full QMC simulations have only been applied to diagonal spin-boson couplings
  using the worm algorithm \cite{PhysRevLett.113.260403}.
  So far, we are still lacking efficient QMC updating schemes for more
  generic spin-boson interactions
  that work on an equal footing for impurity and lattice models.

  Quantum impurity models are at the core of dynamical mean-field theory
  where the coupling to an infinite bath mimics the interaction effects with
  neighboring particles on a lattice \cite{RevModPhys.68.13}. The need
  for efficient impurity solvers
  has motivated the development of
  continuous-time QMC methods for fermions
  where the coupling to the bath is treated in terms of a retarded interaction
  \cite{PhysRevB.72.035122, PhysRevLett.97.076405, RevModPhys.83.349}.
  The most prominent of these methods is based on a hybridization expansion
  and samples fermionic world-line configurations \cite{PhysRevLett.97.076405}.
  The method avoids the negative sign problem by combining the fermionic
   bath propagators into a determinant which only allows for local
   Monte Carlo updates and eventually leads to a cubic scaling in inverse temperature.
   Here, worm sampling is only used
   to obtain improved estimators
   for the Green's function \cite{PhysRevB.92.155102}.
   The method has been extended to include bosonic baths \cite{PhysRevLett.99.146404}
   to solve bosonic impurity problems \cite{Anders_2011},
   spin-boson models \cite{PhysRevB.87.125102}, or Bose-Fermi Kondo models
   \cite{PhysRevB.87.125102,2011JPhCS.273a2050P,PhysRevB.88.245111,PhysRevB.100.014439},
   but the Monte Carlo sampling remained restricted to local updates, even
   in the absence of the fermionic determinant in pure spin-boson models 
   \cite{PhysRevB.87.125102,PhysRevB.100.014439}.
   The self-consistent solution of spin-boson impurity models, \eg, 
   plays an important role for spin glasses
   \cite{Bray_1980, PhysRevLett.70.3339, PhysRevLett.80.389, PhysRevLett.85.840}
   and extensions of dynamical mean-field theory to spin systems \cite{PhysRevB.88.024427}.
   Bosonic modes
   have also been included in the interaction-expansion QMC method
   which can be applied to fermion-boson models on finite lattices
   \cite{PhysRevB.76.035116, PhysRevLett.109.116407, PhysRevB.98.085405}.
   
  In this paper, we introduce a continuous-time QMC method for dissipative
  spin models that combines the advantages of impurity solvers
  with the global worm/directed-loop updates developed
  for lattice models. Our method
  is based on
  the stochastic series expansion (SSE) \cite{PhysRevB.43.5950} 
  with global directed-loop updates \cite{PhysRevE.66.046701}
  and their recent
  generalization to retarded interactions \cite{PhysRevLett.119.097401}.
  In contrast to phonon-coupled systems, the spin-boson
  models considered below do not conserve the total $\spinz$ quantum number,
  which inhibits the local construction of the worm/directed-loop updates.
  To overcome this difficulty, we introduce the new \textit{wormhole update}:
  the two operators of the nonlocal interaction in imaginary time
  can act as sources and sinks for the worm/directed loop
  and therefore allow for nonlocal moves through a world-line configuration.
  It turns out that the rules for constructing the directed loops
  are equivalent to regular spin models with nearest-neighbor interactions---the lattice-site indices
  are just replaced with imaginary-time variables---which allows for an easy transfer
  of the methodology developed in Ref.~\onlinecite{PhysRevE.66.046701}.
  The time dependence of the retarded interaction does not affect the construction
  of the loops because it is sampled during the diagonal updates \cite{PhysRevLett.119.097401};
  the bath propagator can be sampled efficiently for any spectral distribution of the bath modes.
  In the following, we formulate our QMC method for impurity models but
  the wormhole updates can be trivially extended to spin-boson models on a
  finite lattice, as we will discuss at the end of our paper.
  We demonstrate the efficiency of our QMC method for the two-bath spin-boson model.
  Our method reaches significantly lower temperatures than previous QMC approaches
  \cite{PhysRevB.87.125102,PhysRevB.100.014439},
  which allows us to get a precise estimate of the quantum
  critical coupling between the critical and the localized phase that is
  in excellent agreement with previous results of an MPS based approach \cite{PhysRevLett.108.160401}.
  In future, our QMC method will enable efficient simulations of a variety of quantum impurity and lattice models
  coupled to bosonic modes like photons, phonons, magnons, etc., as they appear, \eg,
  in quantum optics, trapped-ion simulators, or solid-state systems.

  Our paper is organized as follows. In Sec.~\ref{Sec:Models}, we
  define the spin-boson models, in Sec.~\ref{Sec:InteractionPicture},
  we show how retarded interactions can be derived from the interaction picture,
  in Sec.~\ref{Sec:Method}, we introduce our QMC method,
  in Sec.~\ref{Sec:Results}, we present our results,
  in Sec.~\ref{Sec:FurtherInteractions}, we discuss extensions of our method, and
  in Sec.~\ref{Sec:Conclusions}, we conclude.

\section{Spin-boson models\label{Sec:Models}}

We consider a generic spin-boson model of the form
\begin{align}
\label{Eq:Ham_gen}
\hat{H} = \hat{H}_\mathrm{s} + \hat{H}_\mathrm{b} + \hat{H}_\mathrm{sb}
\end{align}
that can be split into a spin part $\hat{H}_\mathrm{s}$, a bosonic part $\hat{H}_\mathrm{b}$,
and a spin-boson interaction $\hat{H}_\mathrm{sb}$.
For now, $\hat{H}_\mathrm{s}$ describes a local spin-$1/2$ degree of freedom
$\spinc{\ell}= \frac{1}{2} \hat{\sigma}_\ell$ where $\hat{\sigma}_\ell$, $\ell \in \{x,y,z\}$, are the Pauli matrices (we set $\hbar = 1$).
The bosonic bath is given by a sum of harmonic oscillators,
\begin{align}
\hat{H}_\mathrm{b}
=
\sum_\mu \omega_\mu \bcr{\mu} \ban{\mu} \, ,
\end{align}
where $\bcr{\mu}$ ($\ban{\mu}$) creates (annihilates) a boson in a state $\ket{\mu}$ with frequency $\omega_\mu$. Below, the superindex $\mu$ will refer to a continuum of oscillators and
different components of the bath modes, but it could also label the lattice sites of a finite system.
The spin-boson interaction
\begin{align}
\hat{H}_\mathrm{sb}
=
\sum_\mu \left( \bcr{\mu} \hat{\varrho}^{\phantom{\dagger}}_\mu + \hat{\varrho}_\mu^\dagger \ban{\mu} \right)
\end{align}
couples the bath to a spin operator $\hat{\varrho}_\mu[\spinc{\alpha}]$ that
is model dependent and also contains the spin-boson coupling constant.
In the following, we define two types of
spin-boson models
that fit into the generic form.

The original \textit{spin-boson model} is given by a two-level system which is coupled
to a bosonic bath 
via the $z$ component of the spin, \ie,
\begin{align}
\label{Eq:Ham_spin-boson}
\hat{H}
	=
	- h_x  \spinc{x}
	+ \sum_{q} \omega_{q} \bcr{q} \ban{q}
	+ \sum_{q} \gamma_{q} \left(\bcr{q} + \ban{q} \right) \spinz \, .
\end{align}
The magnetic field $h_x$ induces a finite tunneling between the two levels,
whereas the coupling to a continuous bath with a mode-dependent coupling
constant $\gamma_q$ will localize the spin.
The spin-boson model has been generalized to an interaction with up to three baths, \ie,
\begin{align}
\label{Eq:HamXYZ}
\hat{H}
	=
	- \sum_\ell h_\ell \spinc{\ell}
	+ \sum_{q\ell} \omega_{q} \bcr{q\ell} \ban{q\ell}
	+ \sum_{q\ell} \gamma_{q\ell} \left(\bcr{q\ell} + \ban{q\ell} \right) \spinc{\ell} \, ,
\end{align}
where each bath couples to a different spin component $\ell$.
We set $\gamma_{qx} = \gamma_{qy}$ and refer to this model as
the \textit{XXZ spin-boson model} due to its similarity
to the XXZ model (this will become apparent when we introduce the retarded interactions further below).
For $\gamma_{qz}=0$ we recover the \textit{two-bath spin-boson model} and for
$\gamma_{qx} = \gamma_{qy}= \gamma_{qz}$ the \textit{Bose Kondo model}.
In the absence of the magnetic field, the former model has a U(1) symmetry whereas
the latter is SU(2) symmetric. 
Further details on the U(1)-symmetric case will be discussed in Sec.~\ref{Sec:Results}.
A review of impurity properties can be found in
Ref.~\onlinecite{doi:10.1080/14786430500070396}.

Another Hamiltonian that is often studied in quantum optics
is the \textit{Jaynes-Cummings model} \cite{1443594}
\begin{align}
\hat{H}
	=
	- h_z \spinz
	+
	\sum_{q} \omega_{q} \bcr{q} \ban{q}
	+
	\sum_{q} \gamma_q \left( \bcr{q} \, \spinm + \spinp \, \ban{q}  \right)	\, ,
\end{align}
where the quantized electromagnetic field interacts with a two-level system via
the spin-flip operators $\spinpm = \spinc{x} \pm \im \spinc{y}$.

We consider a coupling to a continuum of bath modes.
The physical properties of the impurity are fully determined by the spectral functions for each bath,
\begin{align}
\label{Eq:Jbath}
J_\ell(\omega)
	=
	\pi \sum_{q} \gamma_{q\ell}^2 \, \delta(\omega - \omega_{q\ell}) \, .
\end{align}
We assume that $J_\ell(\omega)$ has a power-law form with exponent $s$,
\begin{align}
\label{Eq:Jbath_powerlaw}
J_\ell(\omega)
	=
	2\pi \alpha_\ell \, \omega_{\mathrm{c}}^{1-s} \omega^s
	\, ,
	\qquad 0< \omega < \omega_{\mathrm{c}} \, ,
\end{align}
where $s=1$ corresponds to an ohmic bath and $0<s<1$ to the sub-ohmic regime.
Here, $\alpha_\ell$ is the spin-bath coupling constant which we use in the following
and we choose the frequency cutoff $\omega_\mathrm{c}=1$ as the unit of energy.

\section{Interaction picture and retarded interactions\label{Sec:InteractionPicture}}

The spin-boson models introduced in the previous section are quadratic
in the bosonic operators, such that the bath can be integrated out exactly.
Typically, we use the path-integral formalism to derive a retarded interaction
in the system's degrees of freedom. To avoid the intricacies of the spin path integral,
we will show how retarded interactions arise from a perturbation expansion
in the interaction picture. In this way, we will see how the difficulties of previous QMC
approaches are overcome that directly simulate the bosons.
Our formalism is very similar to the hybridization expansion used for fermionic impurity models
\cite{PhysRevLett.97.076405, RevModPhys.83.349}.

We split the Hamiltonian of the full system, $\hat{H} = \hat{H}_0 + \hat{V}$,
into the unperturbed part $\hat{H}_0$ and the perturbation $\hat{V}$.
The Dyson expansion of the partition function reads
\begin{align}
\nonumber
Z
	=
	 \sum_{m=0}^\infty & (-1)^m
	\int_0^\beta d\tau_1  \int_{0}^{\tau_1} d\tau_2  \, {\dots} \int_{0}^{\tau_{m-1}} d\tau_m  \\
	&\times \Tr \big[ e^{-\beta \hat{H}_0}  \hat{V}(\tau_1) \, \hat{V}(\tau_2) \dots  \hat{V}(\tau_m) \big]
\, ,
\label{Eq:Dyson}
\end{align}
where all operators $\hat{V}(\tau) = e^{\tau \hat{H}_0} \hat{V} e^{-\tau \hat{H}_0}$
are ordered according to their imaginary-time variable, \ie,
$\beta \geq \tau_1 \geq \tau_2 \geq \dots  \geq \tau_m \geq 0$
($\beta = 1/T$ is the inverse temperature; we set $\kB = 1$).
For our derivation, it is  convenient to introduce the time-ordering operator $\hat{\mathcal{T}}_\tau$
and rewrite Eq.~(\ref{Eq:Dyson}) in the form
\begin{align}
\nonumber
Z
	=
	 \sum_{m=0}^\infty & \frac{(-1)^m}{m!}
	\int_0^\beta d\tau_1 \int_0^\beta d\tau_2 \, {\dots} \int_{0}^\beta d\tau_m  \\
	&\times \Tr \big[ e^{-\beta \hat{H}_0}  \hat{\mathcal{T}}_\tau \, \hat{V}(\tau_1) \, \hat{V}(\tau_2)  \dots  \hat{V}(\tau_m) \big]
\, .
\label{Eq:DysonTO}
\end{align}
For the generic spin-boson model in Eq.~(\ref{Eq:Ham_gen}), we identify
\begin{align}
\label{Eq:Ham_choice}
\hat{H}_0
\equiv
\hat{H}_\mathrm{b}
\, , \qquad
\hat{V}
\equiv
\hat{H}_\mathrm{sb}
=
\sum_{\mu c} \bop{\mu}{c} \hat{\varrho}^{\bar{c}}_\mu
 \, ,
\end{align}
similar to the hybridization expansion used for fermionic impurity models \cite{RevModPhys.83.349}.
For our derivation of the retarded spin interaction,
we set $\hat{H}_\mathrm{s} = 0$ to simplify the notation.
Later, we can include $\hat{H}_\mathrm{s}$ in $\hat{V}$ without loss of generality.
To distinguish between regular and adjoint operators in Eq.~(\ref{Eq:Ham_choice}),
we introduced the superscript $c$ and its opposite $\bar{c}$.
With this, the interaction expansion becomes
\begin{align}
\nonumber
Z
	=
	 &\sum_{m=0}^\infty  \frac{(-1)^m}{m!}
	\int_0^\beta d\tau_1 \, {\dots} \int_{0}^\beta d\tau_m
	\sum_{\mu_1 \dots \mu_m} \sum_{c_1 \dots c_m} \\ \nonumber
	&\times \Tr_\mathrm{b} \big[ e^{-\beta \hat{H}_\mathrm{b}}  \hat{\mathcal{T}}_\tau \, \bop{\mu_1}{c_1}(\tau_1) \dots \bop{\mu_m}{c_m}(\tau_m) \big] \\
	&\times \Tr_\mathrm{s} \big[ \hat{\mathcal{T}}_\tau \, \hat{\varrho}^{\bar{c}_1}_{\mu_1}(\tau_1) \dots \hat{\varrho}^{\bar{c}_m}_{\mu_m}(\tau_m) \big] 
\, .
\label{Eq:expansion_sb}
\end{align}
The trace splits into separate contributions for spins and bosons.
Because the boson particle-number is conserved,
$\expvbtext{\bullet} = Z_\mathrm{b}^{-1} \Tr_\mathrm{b} [ e^{-\beta \hat{H}_\mathrm{b}} \, \bullet ]$
only gives a nonzero contribution if $m=2n$ and the number of creation and annihilation operators is the same.
The bosonic trace can be further simplified using Wick's theorem. For example, consider the expectation value with
operators ordered as follows:
\begin{align}
\nonumber
& \expvb{  \hat{\mathcal{T}}_\tau \, 
\ban{\mu_{1}}(\tau_{1}) \dots    \ban{\mu_{n}}(\tau_{n})   \,
\bcr{\mu_{n+1}}(\tau_{n+1}) \dots \bcr{\mu_{2n}}(\tau_{2n})
}
\\
&\qquad=
\sum_{\pi \in \mathcal{S}_n}
\prod_{k=1}^n 
D(\omega_{\mu_k},\tau_k - \tau_{n+\pi[k]}) \, \delta_{\mu_k,\mu_{n+\pi[k]}}
\label{Eq:Wickboson}
\, .
\end{align}
Here, $\mathcal{S}_n$ is the symmetric group of order $n$ and $\pi \in \mathcal{S}_n$ is a permutation of $n$ objects.
We introduced the free-boson propagator 
$
D(\omega_\mu,\tau - \tau')
	=
	\expvbtext{ \hat{\mathcal{T}}_\tau \,   \ban{\mu}(\tau) \,  \bcr{\mu}(\tau')  } 
$
which is given by
\begin{align}
\label{Eq:BosonProp}
D(\omega,\tau)
	= \frac{e^{- \omega \tau }}{1-e^{-\beta \omega }} \, ,
	\qquad 0 \leq \tau < \beta \, ,
\end{align}
and fulfills $D(\omega,\tau+\beta) = D(\omega,\tau)$.
The left-hand side of Eq.~(\ref{Eq:Wickboson}) represents only one possible combination of choosing $\{c_1, \dots, c_n\}$
such that we obtain the same number of creation and annihilation operators. In total, there are 
$\binom{2n}{n}$ combinations.
Inserting Eq.~(\ref{Eq:Wickboson}) into Eq.~(\ref{Eq:expansion_sb}), we obtain (we define $\tau_k' = \tau_{n+k}$)
\begin{align}
\label{Eq:expansion_full}
\frac{Z}{Z_\mathrm{b}}
&=
\sum_{n=0}^\infty 
\frac{1}{n!^2} 
\iint_0^\beta d\tau_1 \, d\tau_1'  \sum_{\mu_1} \dots \, \iint_0^\beta d\tau_n \, d\tau_n' \sum_{\mu_n} \\ 
\nonumber
&\times 
\sum_{\pi \in \mathcal{S}_n}
D(\omega_{\mu_1},\tau_1 - \tau_{\pi[1]}') \dots D(\omega_{\mu_n},\tau_n - \tau_{\pi[n]}') \\
&\times \Tr_\mathrm{s} \left[
\hat{\mathcal{T}}_\tau \, \varrhocr{\mu_1}(\tau_1) \, \varrhoan{\mu_1}(\tau_{\pi[1]}') \dots  \varrhocr{\mu_n}(\tau_n) \, \varrhoan{\mu_n}(\tau_{\pi[n]}')
\right]
 \, .
\nonumber
\end{align}
The sum over all permutations can be evaluated by relabeling the variables, which gives another factor
of $n!$ in Eq.~(\ref{Eq:expansion_full}).
Eventually, the perturbation expansion can be recast as
\begin{align}
\label{Eq:expansion_full_final}
\frac{Z}{Z_\mathrm{b}}
=
\sum_{n=0}^\infty 
\frac{1}{n!} 
 \Tr_\mathrm{s} \Bigg\{ \hat{\mathcal{T}}_\tau
 \bigg[
&\iint_0^\beta d\tau \, d\tau'  \sum_{\mu} 
\varrhocr{\mu}(\tau) \\
&\quad\times  D(\omega_\mu,\tau-\tau') \, \varrhoan{\mu}(\tau')
\bigg]^n
\Bigg\} \, .
\nonumber
\end{align}
After reexponentiation of the resulting interaction term in Eq.~(\ref{Eq:expansion_full_final}),
the partition function becomes
\begin{align}
\label{Eq:Zret}
Z =
Z_\mathrm{b}
\Tr_\mathrm{s} \hat{\mathcal{T}}_\tau \, e^{-\S_\mathrm{ret}} \, .
\end{align}
As a result, the spin-boson coupling has generated a retarded spin-spin interaction
\begin{align}
\label{Eq:Sret_gen}
\S_\mathrm{ret}
	=
	- \iint_0^\beta d\tau d\tau' \sum_\mu
	\varrhocr{\mu}(\tau) \, D(\omega_\mu,\tau-\tau') \, \varrhoan{\mu}(\tau')
\end{align}
that is mediated by the free-boson propagator (\ref{Eq:BosonProp}).
Note that for our choice of $\hat{H}_0$ the time evolution of the spins is trivial because
$\hat{H}_0$ does not include any spin operators.
However, we still need the $\tau$ labels for the time ordering of the spins in the perturbation expansion
as well as for the time dependence of the boson propagators.

Finally, we want to specify the retarded spin interactions for the spin-boson models introduced
in the previous section. For the XXZ spin-boson model we get
\begin{align}
\nonumber
\S_\mathrm{ret}
	&=
	- \frac{1}{\pi} \sum_\ell \int_0^\infty d\omega \, J_\ell(\omega)  \iint_0^\beta d\tau d\tau'  \\
	&\qquad \qquad \qquad \times  \spinc{\ell}(\tau) \, D_+(\omega,\tau-\tau') \, \spinc{\ell}(\tau') \, ,
\label{Eq:SretXXZ}
\end{align}
whereas the Jaynes-Cummings Hamiltonian leads to
\begin{align}
\nonumber
\S_\mathrm{ret}
	&=
	- \frac{1}{\pi} \int_0^\infty d\omega \, J(\omega)  \iint_0^\beta d\tau d\tau' \\
	&\qquad \qquad \qquad \times \spinp(\tau) \, D(\omega,\tau-\tau') \, \spinm(\tau') \, .
\label{Eq:SretJC}
\end{align}
Here, we have already taken the continuum limit for the bosonic bath.
We explicitly see that
the spectral function defined in Eq.~(\ref{Eq:Jbath}) completely determines
the effect of the bath on the spin subsystem.
For the power-law spectrum in Eq.~(\ref{Eq:Jbath_powerlaw}),
the frequency average of $D(\omega,\tau)$ induces a long-range interaction 
in imaginary time that vanishes as $1/\tau^{s+1}$ for $\omega_\mathrm{c} \tau \gg 1$.
Because the retarded interaction for the XXZ spin-boson model is symmetric
under $\tau \leftrightarrow \tau'$, we have replaced $D(\omega,\tau)$ with
the symmetrized propagator $D_+(\omega,\tau) = [D(\omega,\tau)+D(\omega,\beta-\tau)]/2$
in Eq.~(\ref{Eq:SretXXZ}).

\section{Quantum Monte Carlo method\label{Sec:Method}}

\subsection{Basic ideas of the Monte Carlo method\label{Sec:MethodBasics}}

Before we describe the mathematical formalism of our QMC method, 
we first want to introduce the main ideas that underlie our approach.
We use
the framework of the directed-loop algorithm \cite{PhysRevE.66.046701}
which was originally formulated in the SSE representation \cite{PhysRevB.43.5950}
and recently generalized  to retarded interactions \cite{PhysRevLett.119.097401}.
We have shown in the previous section that a generic spin-boson model of the form
of Eq.~(\ref{Eq:Ham_gen}) can be mapped to a spin problem with a retarded interaction.
In this way, we can avoid the difficulties of direct boson sampling which often occurs
when the particle number is not conserved.
As the resulting retarded spin
interaction in Eq.~(\ref{Eq:Sret_gen}) does not conserve the total $\spinz{}$, either, 
we will introduce the nonlocal \textit{wormhole update} to provide an efficient sampling
using the directed-loop algorithm.

As a starting point for the formulation of our method, we consider
the partition function in Eq.~(\ref{Eq:Zret}), where the bosonic
bath has been integrated out to obtain a retarded spin interaction.
%In analogy to the SSE representation,
Following Ref.~\onlinecite{PhysRevLett.119.097401},
we expand the time-ordered exponential in the full interaction $\S_\mathrm{ret}$,
\ie, we perform an interaction expansion around the trivial spin part $\S_0 = 0$,
to obtain Eq.~(\ref{Eq:expansion_full_final}). Such an expansion in the full Hamiltonian is
\textit{the} characteristic feature that leads to the SSE representation, even if we start from the interaction
representation.
Then, the time evolution of the spin operators $\spinc{\ell}(\tau)$ becomes trivial and their $\tau$ labels
are only required to perform the time ordering.
For equal-time (\ie, time-independent) Hamiltonians, the imaginary-time integrals in the resulting expansion
can be calculated exactly, leading to an exact mapping to the SSE representation \cite{PhysRevB.56.14510,Assaad2008},
which does not have an explicit time dependence.
In the presence of retarded interactions, which is the focus of this paper, imaginary-time variables have to be
sampled explicitly in our Monte Carlo scheme to take the time-dependence of the boson propagator correctly into account. 
It was shown in Ref.~\onlinecite{PhysRevLett.119.097401} that the exact sampling of the continuous times
only enters during the diagonal updates (as discussed below), so that the remaining parts of the directed-loop algorithm
can be formulated as in the SSE representation.

\begin{figure}
  \includegraphics[width=0.75\linewidth]{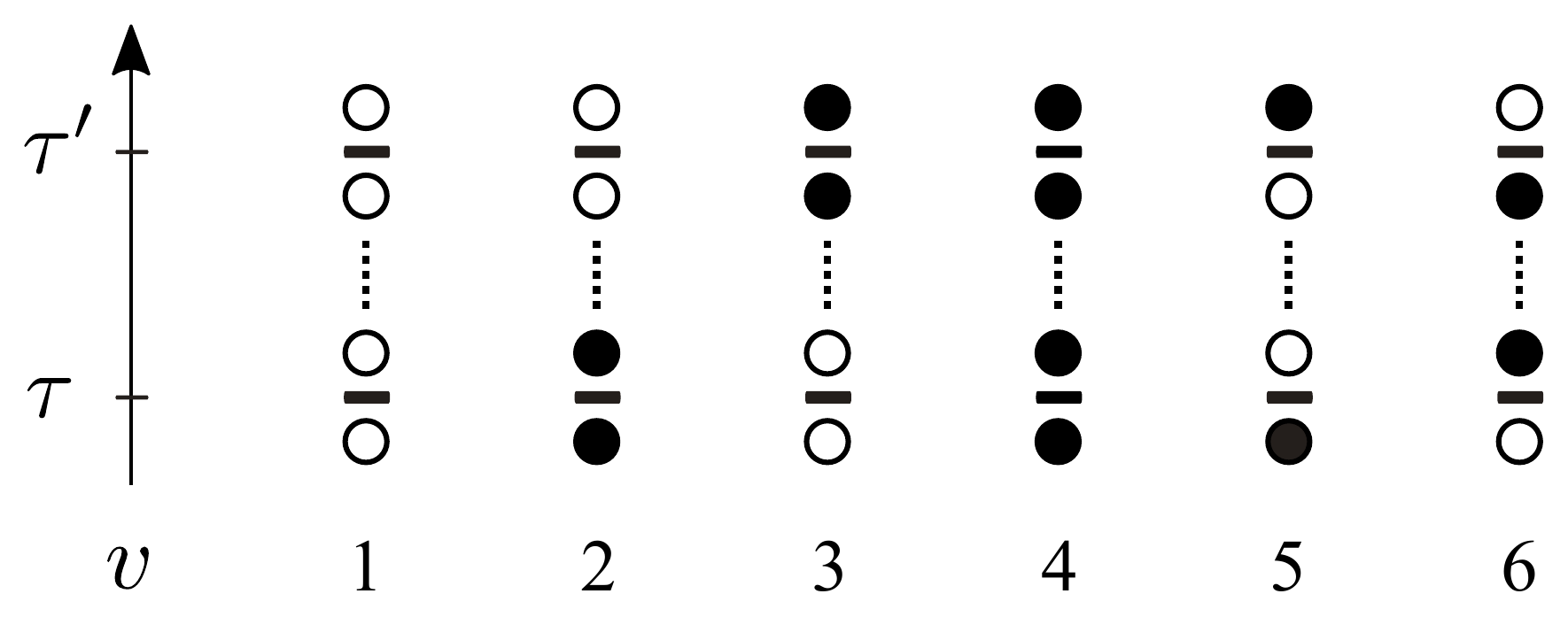}
  \caption{\label{fig:vertextypes}%
Graphical representation of the retarded interaction vertex.
The black bars at times $\tau$ and $\tau'$ correspond to the two operators of a vertex, whereas filled (open) circles illustrate the
spin states with $\spinz$ eigenvalue $1/2$ ($-1/2$) before and after an operator is applied.
The two operators are connected by a dashed line representing the free-boson propagator.
Diagonal vertices are identified by vertex types $v=1$--$4$, whereas  the off-diagonal terms $\spinm(\tau) \, \spinp(\tau')$
and $\spinp(\tau) \, \spinm(\tau')$ are labeled by $v=5$ and $v=6$, respectively.
  }
\end{figure}

\begin{figure}
  \includegraphics[width=0.9\linewidth]{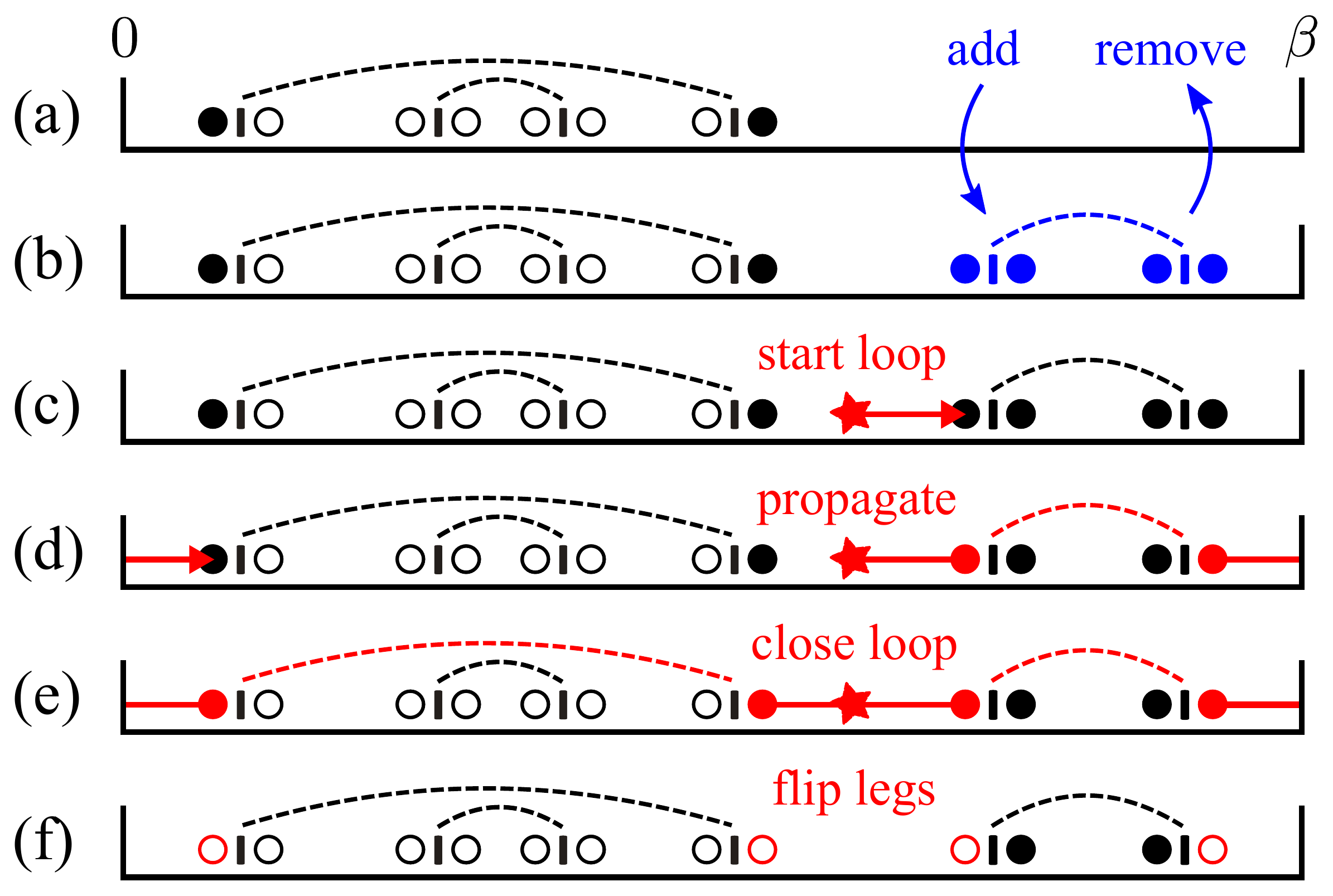}
\caption{\label{fig:scetch}%
(a) Illustration of a world-line configuration with two retarded vertices.
Imaginary-time propagation of the initial state $\ket{\alpha}$ is from left to right.
Because the propagated state can only change when an operator is applied, it is sufficient
to denote the spin configuration at the legs of each vertex. The addition or removal
of a diagonal vertex between (a) and (b) does not change the world-line configuration.
(c)~To initialize the directed-loop updates, we pick a random starting time
and propagate the head of the loop until it reaches the entrance leg of a vertex. (d)~We choose
an exit leg according to the probabilities determined by the directed-loop equations. Here,
the loop head uses the nonlocal connection established by the free-boson propagator to perform a wormhole move.
(e)~This procedure is repeated until the loop head returns to its starting point. (f)~Eventually,
the world-line configuration is updated, \ie, 
we flip all legs connected by the loop and in our case also the initial state. Thereby, diagonal vertices
can be transformed into off-diagonal ones and vice versa.
  }
\end{figure}

We first want to discuss how the trace over the spin operators is related to the original SSE formulation.
%Then, the $\tau$ labels
%of the spin operators are only required to perform the time ordering.
The time-ordered product
of spin operators corresponds to the operator sequence in the SSE representation.
From our derivation in Sec.~\ref{Sec:InteractionPicture}, it is clear that the time ordering does not
lead to any negative signs. The trace over the spins,
$\Tr_\mathrm{s} \bullet = \sum_\alpha \braket{\alpha | \bullet | \alpha}$,
runs over the initial state $\ket{\alpha}$ which is usually chosen in the $\spinz{}$ basis. We assume that the operators
in $\S_\mathrm{ret}$ are non-branching, \ie,
$\hat{\varrho}_{\mu_p}^{\bar{c}_p}(\tau_p) \ket{\alpha_{p}} \sim \ket{\alpha_{p-1}}$
uniquely propagates a basis state to another basis state;
here we define the time ordering as in Eq.~(\ref{Eq:Dyson}) (with a decreasing propagation index) and set
$\ket{\alpha_0} = \ket{\alpha_{2n}} = \ket{\alpha}$.
The imaginary-time evolution of the initial state can be visualized
using a world-line picture (note that a world line is defined by the time dependence of the propagated spin state and not by the operators).
Because the propagated state $\ket{\alpha_p}$ only changes when an operator is applied,
it is sufficient to find a graphical representation of the interaction vertex.
The possible vertex types for $\S_\mathrm{ret}$ are illustrated in Fig.~\ref{fig:vertextypes}.
Each vertex consists of two operators, the boson propagator, and the spin states before and
after the operators are applied, which are represented by two black bars, a dashed line,
and four legs, respectively.
For our spin-$1/2$ models, the legs are illustrated as filled (open) circles and  correspond to $\spinz{}$ eigenstates
with eigenvalue $1/2$ ($-1/2$).
In total, 
there are six vertex types for the retarded interactions in Eqs.~(\ref{Eq:SretXXZ}) and (\ref{Eq:SretJC}): four \textit{diagonal} vertices which leave the world-line
configuration unchanged and two \textit{off-diagonal} vertices where the spin states are flipped.
The latter correspond to $\spinm(\tau) \, \spinp(\tau')$ and $\spinp(\tau) \, \spinm(\tau')$.
A typical world-line configuration that includes both types of vertices is shown in Fig.~\ref{fig:scetch}(a).
For a more formal definition of the interaction vertices and the configuration space of our QMC method see Sec.~\ref{Sec:Vertices}.

We use Markov chain Monte Carlo sampling to evaluate the sum over all world-line configurations stochastically.
To formulate an ergodic algorithm, we need two types of updates: the diagonal updates and the
directed-loop updates.

The main purpose of the \textit{diagonal updates} is to change the average
expansion order, to replace existing vertices with new ones, and---in addition
to the original SSE method---to sample the imaginary-time variables.
Because diagonal operators do not change the world-line configuration,
we propose to add or remove diagonal vertices using the Metropolis
scheme defined in Sec.~\ref{Sec:diagonalupdates}.
An example for the addition or the removal of such a 
 vertex is given in Figs.~\ref{fig:scetch}(a) and \ref{fig:scetch}(b).
The acceptance rates for adding a new vertex can be significantly improved
by sampling the time difference between the vertex operators according to the free-boson propagator.
A block of diagonal updates requires $\mathcal{O}(n)$ proposals
in order to replace
an extensive number of
vertex variables with new ones. 

The efficiency of our Monte Carlo method relies on the \textit{directed-loop updates} \cite{PhysRevE.66.046701},
a global updating scheme that changes an extensive subset of world-line segments 
along a closed loop. The loop is constructed as follows:
First, we pick a random starting time
to insert a pair of spin-flip operators.
One of these operators
is identified as the head of the loop
that can move through imaginary time as a world-line discontinuity, whereas the other
operator becomes the tail of the loop that remains fixed at the starting time.
Because the time evolution for the spin operators is trivial ($\hat{H}_0$ does not contain any spin terms),
the head will proceed until it reaches the leg of an operator that is part of a vertex. In the conventional
directed-loop update, we would choose an exit leg of a local vertex with a probability determined by
the directed-loop equations, a local version of detailed balance. It turns out that this is still true for the nonlocal vertex
of the retarded interaction. As a result, the loop head entering the vertex in Fig.~\ref{fig:scetch}(c) via the left leg of the left operator
can either bounce and reverse its direction of propagation, continue straight through the left operator, 
or exit at the right operator. In Fig.~\ref{fig:scetch}(d), we selected the last option which we call a
\textit{wormhole update} because the loop head uses the connection established by the free-boson propagator
as a wormhole to instantly tunnel between well-separated points in imaginary time---the name is also chosen in analogy to the worm algorithm where the loop is called a worm \cite{Prokofev:1998aa}.
Once the loop exits the leg of a vertex, it will propagate to the entrance leg of the next vertex and the same procedure will apply again until the loop head returns to its tail, as depicted in Fig.~\ref{fig:scetch}(e).
When the loop is closed, we flip all spin configurations along the loop, as shown in Fig.~\ref{fig:scetch}(f).
Thereby, diagonal vertices can be transformed into off-diagonal ones and vice versa.
Because the directed loop is able to touch an extensive number of vertices, it represents
 a global update move. The computational effort to construct the loop scales linearly in the number
of touched vertices. Further details on the construction of the directed loops are given in Sec.~\ref{Sec:WormholeUpdates}.

With the diagonal and directed-loop updates described above, our QMC algorithm samples the entire diagrammatic expansion
of diagonal and off-diagonal operators
and therefore fulfills ergodicity. The diagonal updates sample the expansion order $n$ by adding and removing diagonal vertices at arbitrary
imaginary times, whereas the directed-loop updates transform diagonal operators into off-diagonal ones and vice versa. 
The latter transformation reaches all vertex types with a nonzero weight, as described in the original formulation of
the directed-loop updates \cite{PhysRevE.66.046701}.

Altogether, we combine the framework of retarded interactions---that is commonly used for impurity solvers
to avoid the difficulties of direct boson sampling \cite{RevModPhys.83.349}---with the global directed-loop updates originally developed for lattice models \cite{PhysRevE.66.046701}. 
The local construction of the wormhole updates is possible for bosonic baths with
positive-definite propagators, whereas fermionic baths require the numerically-expensive calculation of a determinant to
avoid the negative-sign problem \cite{PhysRevLett.97.076405}.
The formulation of our algorithm is closely related to the standard
formulation of the directed-loop algorithm \cite{PhysRevE.66.046701}. However, the nonlocality in imaginary time
requires some changes in the implementation which we will discuss in the Appendix.
Although we have developed the wormhole updates in the framework of the directed-loop algorithm,
their underlying ideas can be
transferred to related QMC methods based on worm/directed-loop updates.

\subsection{Configuration space and vertex weights\label{Sec:Vertices}}

A Monte Carlo configuration $\conf = \{ n, \confn, \ket{\alpha} \}$ is defined by the
expansion order $n$, the ordered vertex list $\confn=\{ \nu_1, \dots, \nu_n \}$,
and the initial state $\ket{\alpha}$ in the local $\spinz{}$ basis.
Each vertex variable $\nu = \{\typeint, a, \omega, \tau, \tau' \}$ includes
a frequency $\omega$ and two imaginary-time values $\tau$ and $\tau'$.
The operator type $a$ distinguishes between diagonal and off-diagonal parts of the vertex,
whereas the interaction type $\typeint$ labels different kinds of interactions%
---the latter is only relevant if we combine the spin-boson models with additional interactions,
as discussed in Sec.~\ref{Sec:FurtherInteractions}.
The interaction vertex for our spin-boson models is given by
\begin{align}
\nonumber
\S
	&=
	- \int_0^\infty d\omega \, \mathcal{J}(\omega) \iint_0^\beta d\tau d\tau' \sum_a 
	\\
	 &\qquad \qquad \qquad \times  P(\omega,\tau-\tau') \, \hat{h}_a(\tau,\tau') \, .
\end{align}
Here, we have rescaled the free-boson propagator and the spectral density,
\begin{align}
P(\omega,\tau) = \omega \, D(\omega,\tau) \, , \quad
\mathcal{J}(\omega) = \frac{J(\omega) / \omega}{\int d\omega \, J(\omega) / \omega} \, ,
\end{align}
such that the former becomes a probability distribution in $\tau$ and the latter
in $\omega$.
For the power-law spectrum in Eq.~(\ref{Eq:Jbath_powerlaw}), we obtain
$\mathcal{J}(\omega)
	=
	s \, \omega_\mathrm{c}^{-s} \omega^{s-1}$.
The normalization factor for the spectral density can be absorbed into the operator
part $\hat{h}_a(\tau,\tau')$.

Our Monte Carlo method samples the perturbation expansion of the partition
function, $Z = Z_\mathrm{b} \sum_{\conf} W(\conf)$, as in Eq.~(\ref{Eq:expansion_full_final}). The total weight
of a configuration is given by
\begin{gather}
\label{Eq:MC_weight}
W(\conf)
	=
	\frac{1}{n!} \prod_{p=1}^n \mathcal{W}_{\nu_p}
	\, , \qquad \text{where}
	\\
\mathcal{W}_\nu
	=
	\mathcal{J}(\omega) \, P(\omega,\tau-\tau') \, W[\hat{h}_a(\tau,\tau')] \, d\omega \, d\tau \, d\tau'
\label{Eq:vertex_weight}
\end{gather}
is the weight of a single vertex. The expectation value of the operator part
$W[\hat{h}_a(\tau,\tau')]$ is fully determined by the weight $W_v$ of one of the six
vertex types $v$ in Fig.~\ref{fig:vertextypes}.

For the XXZ spin-boson model, the off-diagonal operator part is given by
\begin{align}
\hat{h}_1(\tau,\tau')
\label{Eq:h1XXZ}
	=
	 \frac{\lambda_{xy}}{2} \left[ \spinp(\tau) \, \spinm(\tau') + \spinm(\tau) \, \spinp(\tau') \right] \, ,
\end{align}
whereas the diagonal part becomes
\begin{align}
\label{Eq:h2XXZ}
\hat{h}_2(\tau,\tau')
	=
	C &+ \lambda_z \, \spinz(\tau) \, \spinz(\tau')  \nonumber \\
	 &+ \frac{h_z}{2} \left[\spinz(\tau) + \spinz(\tau')\right] \, .
\end{align}
Here, we have defined the couplings $\lambda_\ell = 2 \alpha_\ell \omega_\mathrm{c} / s$
which include the normalization of the spectral function. Note that we have dropped
unit operators at times $\tau$ and $\tau'$, \eg, $C$ corresponds to
$C \, \hat{\mathbb{1}}(\tau) \, \hat{\mathbb{1}}(\tau')$.
We can include the magnetic field $h_z$ in the retarded interaction because
$\int d\omega \, \mathcal{J}(\omega) \int d\tau' \, P(\omega,\tau-\tau') = 1$.
We have added a constant $C$ to $\hat{h}_2(\tau,\tau')$
to obtain positive weights. For the different vertex types $v$ defined in Fig.~\ref{fig:vertextypes}, the weights are given by
\begin{align}
\nonumber
W_1 &= C + \frac{\lambda_z}{4} - \frac{h_z}{2} \, ,
\qquad
W_2 = W_3 = C - \frac{\lambda_z}{4} \, ,
\\
W_4 &= C + \frac{\lambda_z}{4} + \frac{h_z}{2} \, ,
\qquad
W_5 = W_6 = \frac{\lambda_{xy}}{2} \, .
\label{Eq:vertexweights_XXZ}
\end{align}
We choose $C \geq \max \big[ \frac{\lambda_z}{4}, \frac{\absolute{h_z}}{2} - \frac{\lambda_z}{4} \big]$
to get positive weights.
The weights $W_v$ for the XXZ spin-boson model are equivalent
to the ones for the ferromagnetic XXZ model with a nearest-neighbor spin exchange \cite{PhysRevE.66.046701}.

For the Jaynes-Cummings model, we have
\begin{align}
\hat{h}_1(\tau,\tau')
	&=
	 \frac{\lambda_{xy}}{2} \, \spinp(\tau) \, \spinm(\tau')  \, ,
	 \\
\hat{h}_2(\tau,\tau')
	&=
	C 
	 + \frac{h_z}{2} \left[\spinz(\tau) + \spinz(\tau')\right] \, .
\end{align}
The vertex weights are similar to the ones in Eq.~(\ref{Eq:vertexweights_XXZ}) but
$\lambda_z = 0$ and $W_5 = 0$.

\subsection{Diagonal updates\label{Sec:diagonalupdates}}

The diagonal updates involve adding or removing a single vertex $\hat{h}_2(\tau,\tau')$
using the Metropolis-Hastings algorithm. The Metropolis acceptance probability
\begin{align}
\label{Eq:Metropolis_A}
A(\conf \to \conf') = \min[1, R({\conf \to \conf'})] 
\end{align}
 is determined by the acceptance ratio
\begin{align}
R{(\conf \to \conf')} = \frac{ W(\conf') \, T_0(\conf' \to \conf)}{W(\conf) \, T_0(\conf \to \conf')} \, ,
\end{align}
which depends on the Monte Carlo weight $W(\conf)$ defined in Eq.~(\ref{Eq:MC_weight})
and the proposal probabilities $T_0(\conf \to \conf')$ specified in the following.

We propose the addition of a new vertex with probability density
 \begin{align}
 \label{Eq:DiagProposeAdd}
 T_0({\conf_n\to \conf_{n+1}}) = \frac{\mathcal{J}(\omega) \, P(\omega,\tau-\tau') \, p_{\typeint} \, d\omega \, d\tau \, d\tau' }{\beta \left(n+1\right)} \, ,
 \end{align}
 where only the first time variable is chosen randomly, but the frequency and the second time variable
 are sampled from $\mathcal{J}(\omega)$ and $P(\omega,\tau-\tau')$, as explained below.
 Note that there are $n+1$ possibilities to insert the new vertex into the ordered list $\conf_{n+1}$.
 We included an additional probability $p_{\typeint}$ to pick an interaction type $\typeint$ from a set of interaction terms,
 as further discussed in Sec.~\ref{Sec:FurtherInteractions}; for now, $p_{\typeint}=1$.
 For the removal of a randomly chosen diagonal vertex, we get
 \begin{align}
 T_0({\conf_{n+1}\to \conf_{n}}) = \frac{1}{n_{2}+1} \, ,
 \end{align}
 where $n_2$ is the number of all diagonal vertices in $\conf_n$.
 With the ratio of the Monte Carlo weights in Eq.~(\ref{Eq:MC_weight}), 
$
W_\alpha(\conf_{n+1}) / W_\alpha(\conf_n) = \frac{1}{n+1} \mathcal{J}(\omega) \, P(\omega,\tau-\tau') \,
  W_v \, d\omega \, d\tau \, d\tau' 
$,
we obtain the Metropolis ratios
\begin{align}
R(\conf_n \to \conf_{n+1})
	&=
	\frac{\beta \, W_v}{\left(n_2+1\right) p_{\typeint}} \, , \\
R(\conf_n \to \conf_{n-1})
	&=
	\frac{n_2 \, p_{\typeint}}{\beta \, W_v} 
\end{align}
for the addition and removal of a vertex, respectively.
Because we included $\mathcal{J}(\omega)$ and $P(\omega,\tau-\tau')$
in the proposal probabilities, they drop out of the acceptance rates.
In this way, we ensure high acceptance probabilities for all parameters.

We use inverse transform sampling to draw $\omega$ and $\tau-\tau'$
from the probability distributions $\mathcal{J}(\omega)$ and $P(\omega,\tau-\tau')$, respectively.
Given a probability distribution function $f(x)$, we calculate its
cumulative distribution function $F(x) = \int_0^x dx' f(x')$. If we choose a
uniformly-distributed random number $\xi \in [0,1)$, then $F^{-1}(\xi)$
returns a random number drawn from $f(x)$.
In our case, the cumulative distributions for $\mathcal{J}(\omega)$ and $P(\omega,\tau-\tau')$
can be inverted analytically. We first draw $\omega$ from the power-law spectrum
in Eq.~(\ref{Eq:Jbath_powerlaw}), \ie,
\begin{align}
\omega
	=
	\omega_\mathrm{c} \, (1-\xi)^{1/s} \, .
\end{align}
Afterwards, we use the chosen $\omega$ value to draw
$\tau-\tau'$ from the boson propagator in Eq.~(\ref{Eq:BosonProp}), \ie,
\begin{align}
\label{Eq:tau_sampling}
\tau-\tau'
	=
	- \frac{1}{\omega} \ln[ 1- \xi (1-e^{-\beta\omega}) ] \, .
\end{align}
To draw the time difference from the symmetrized propagator $P_+(\omega,\tau-\tau')$,
we can derive a similar formula or just
replace $\tau-\tau' \to \beta - (\tau-\tau')$ in Eq.~(\ref{Eq:tau_sampling}) with probability $1/2$.
Note that we draw $\omega \in(0,\omega_\mathrm{c}]$ and $\tau-\tau' \in [0,\beta)$.

\subsection{Directed-loop and wormhole updates\label{Sec:WormholeUpdates}}

We have introduced the ideas behind the directed-loop updates and their extension
to wormhole updates already in Sec.~\ref{Sec:MethodBasics}. In the following, we
want to elaborate on their mathematical formulation.

The directed-loop updates use an extended configuration space to connect
two regular Monte Carlo configurations that require an extensive number of local changes.
The corresponding world-line configurations only differ by the presence of a closed loop
along which the spins are flipped. 
To construct this global update, we need to fulfill the detailed-balance
condition. 
It was shown in Ref.~\onlinecite{PhysRevE.66.046701} that this global requirement
reduces to a set of local conditions for each vertex, known as the directed-loop equations, because the Monte Carlo weight in Eq.~(\ref{Eq:MC_weight})
factorizes. The proof in Ref.~\onlinecite{PhysRevE.66.046701} does not make any assumptions on the
structure of the vertex, therefore it is also valid for our retarded interaction. Because
$\mathcal{J}(\omega)$ and $P(\omega,\tau-\tau')$ are global prefactors for each vertex, \ie,
they do not depend on the vertex type in Eq.~(\ref{Eq:vertex_weight}), they drop out
of the directed-loop equations \cite{PhysRevLett.119.097401}. Hence, the latter can be formulated for the vertex weights $W_v$
as for the original method \cite{PhysRevE.66.046701}. In the extended configuration space,
each vertex is assigned an entrance and exit leg for the directed loop and the corresponding
weights become $W_v(l_\mathrm{in},l_\mathrm{out})$.
The directed-loop equations,
\begin{gather}
\label{Eq:dirloop1}
W_v(l_1,l_2) = W_{\bar{v}}(l_2,l_1) \, ,
\\
\sum_{l_2} W_v(l_1,l_2) = W_v \, ,
\label{Eq:dirloop2}
\end{gather}
correspond to a local version of detailed balance and the conservation of probability. Here,
the vertex type $\bar{v}$ is related to $v$ by flipping the spins along the assigned loop segment.
If the directed loop enters a vertex at leg $l_1$, $W_v(l_1,l_2)/W_v$ gives the probability
to exit at leg $l_2$.

\begin{figure}
  \includegraphics[width=0.9\linewidth]{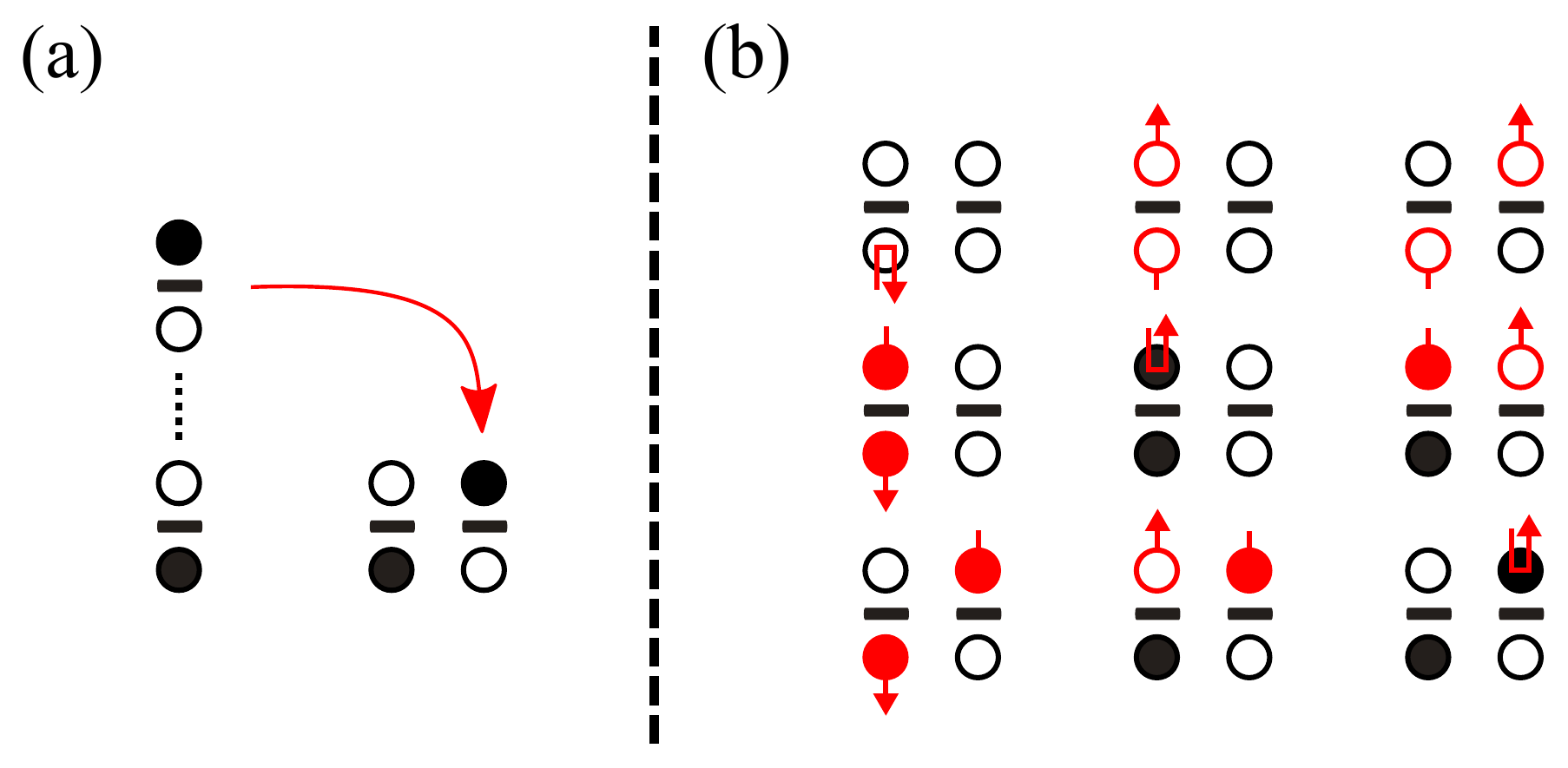}
\caption{\label{fig:asstab}%
(a)~For the directed-loop updates, the retarded vertex of a spin-boson model is equivalent to the nearest-neighbor vertex of a lattice model.
(b) The assignment tables to solve the directed-loop equations are constructed in the same way as for lattice models
(see the main text for further explanations).
Here, we have assigned a directed-loop segment to each vertex and colored those vertex legs in red which get flipped during the directed-loop update.
  }
\end{figure}

Before we explain how to solve the directed-loop equations, we want to take another look
at the vertex structure depicted in Fig.~\ref{fig:vertextypes}. As for regular spin models
on a finite lattice (\eg, Heisenberg or XXZ models), each vertex has four legs.
We have argued before that the nonlocality of our vertex
does not play a role in the directed-loop equations.
We find that the vertices in Fig.~\ref{fig:vertextypes} map to the ones for regular
spin models with nearest-neighbor interactions if we take the subvertex at $\tau'$ and place it to the right
of the subvertex at $\tau$ [as illustrated in Fig.~\ref{fig:asstab}(a)]. Hence, we can use the same techniques as for
regular spin models to solve the directed-loop equations. Moreover,
the vertex weights for the XXZ spin-boson model in Eq.~(\ref{Eq:vertexweights_XXZ})
are equivalent to the ones for the ferromagnetic XXZ spin model on a finite lattice.
As a result, the directed-loop equations have the same solution for both cases. In particular,
loops can be constructed deterministically in the SU(2) symmetric case \cite{PhysRevB.59.R14157}.

For reasons of completion, we give a short outline on how to solve the directed-loop equations analytically
for the spin-boson models considered in this paper.
In more general situations, the directed-loop equations can be solved
using linear programming techniques \cite{PhysRevE.71.036706}.
Consider the assignment table illustrated in Fig.~\ref{fig:asstab}(b).
It is constructed in such a way that each row corresponds to Eq.~(\ref{Eq:dirloop2})
and the off-diagonal elements are related by Eq.~(\ref{Eq:dirloop1}).
The bounce probabilities are given on the diagonal.
For the assignments chosen in Fig.~\ref{fig:asstab}(b), we obtain the set of equations
\begin{align}\nonumber
b_1 + a + b = W_{1} \, , \\
a + b_2 + c = W_{2} \, , \\\nonumber
b + c + b_3 = W_{5} \, ,
\end{align}
which we can solve for $a$, $b$, and $c$ as follows:
\begin{align}\nonumber
a &= \frac{1}{2} \left[ W_{1} + W_{2} - W_{5} - b_1 - b_2 + b_3 \right] \, , \\
b &= \frac{1}{2} \left[ W_{1} - W_{2} + W_{5} - b_1 + b_2 - b_3 \right] \, , \\\nonumber
c &= \frac{1}{2} \left[ -W_{1} + W_{2} + W_{5} + b_1 - b_2 - b_3 \right] \, .
\end{align}
Our guiding principle is to minimize the weights $b_1$, $b_2$, and $b_3$ for the bounce moves
as they reduce the efficiency of the algorithm.
If we plug in the weights for the XXZ spin-boson model, we get an explicit solution,
\begin{align}\nonumber
a &= \frac{1}{2} \left[ 2C - \frac{h_z}{2} -\frac{\lambda_{xy}}{2} - b_1 - b_2 + b_3 \right] \, , \\
b &= \frac{1}{2} \left[ \frac{\lambda_z}{2} - \frac{h_z}{2} + \frac{\lambda_{xy}}{2} - b_1 + b_2 - b_3 \right] \, , \\\nonumber
c &= \frac{1}{2} \left[ -\frac{\lambda_z}{2} + \frac{h_z}{2} + \frac{\lambda_{xy}}{2} + b_1 - b_2 - b_3 \right] \, .
\end{align}
We can always obtain positive solutions if we adjust the constant shift $C$ and one of the bounce
weights $b_1$ or $b_2$. 
In the absence of an external magnetic field $h_z$, we can obtain bounce-free solutions for $\lambda_z \leq \lambda_{xy}$.
For the SU(2) symmetric case, \ie, $\lambda_z = \lambda_{xy}$, we can choose
$C= \lambda_z / 4$ to construct loops where the entrance leg exactly determines the exit leg,
as it is also the case for the Heisenberg model \cite{PhysRevB.59.R14157}.
The directed-loop equations have to be solved for all possible assignment tables.
Similar solutions can be derived for the Jaynes-Cummings interaction.

\subsection{Observables}

The calculation of observables is equivalent to standard world-line QMC methods.
In the following, we only give a short overview over the most common ones.

The properties of the spin system can be accessed from the
imaginary-time correlation functions
\begin{align}
C_\ell (\tau - \tau')
	=
	\expv{\spinc{\ell}(\tau) \, \spinc{\ell}(\tau')}
\end{align}
and the corresponding susceptibilities 
\begin{align}
\chi_\ell(\im \Omega_m)
	=
	\frac{1}{\beta} \iint_0^\beta d\tau d\tau' e^{\im \Omega_m (\tau-\tau')} C_\ell (\tau - \tau') \, .
\end{align}
Here, $\Omega_m = 2\pi m / \beta$, $m\in \mathds{Z}$, are the bosonic Matsubara frequencies.
We also define the static susceptibilities $\chi_\ell = \chi_\ell(\im \Omega_0)$.
The $z$ components of these observables do not change the world-line configurations and can be obtained
directly from the propagated state $\ket{\alpha_p}$, whereas the off-diagonal components
can be accessed by tracking the propagation of the directed loop during the directed-loop updates.
The latter is possible because the loop head and tail are identified with spin-flip operators---depending on the model, it might be more convenient to identify them either with $\hat{\sigma}_x$ operators or with $\spinp$ and $\spinm$ operators.
In contrast to the SSE representation,
we have direct access to the imaginary-time values of each vertex which simplifies the calculation
of time-displaced observables for both cases.
For a detailed discussion of the diagonal and off-diagonal measurements
see Refs.~\onlinecite{PhysRevB.56.14510,PhysRevE.64.066701}.

The properties of the bath cannot be accessed directly from the world-line configurations
because the bosons have been integrated out. However, bosonic observables
can be derived from higher-order spin-spin correlation functions with the help of generating
functionals and eventually be recovered 
from the distribution of vertices \cite{PhysRevB.94.245138}. 
For example,
$\expvtext{n} = - \beta \expvtext{\hat{H}_\mathrm{s} + \hat{H}_\mathrm{sb}}$ relates
the average expansion order to the energy of the spin subsystem.
Estimators for the bosonic energy, the boson propagator, or the specific heat have been
derived in Refs.~\onlinecite{PhysRevB.94.245138,PhysRevB.98.235117} for a single bath frequency.
They can be generalized to a continuous bath, but one has to define an additional mapping
from discrete frequencies to the continuum following Refs.~\onlinecite{Bulla_1997, PhysRevLett.91.170601}
leaving some ambiguity for the bath properties.

\section{Results\label{Sec:Results}}

To demonstrate the efficiency of our QMC method, we consider the quantum phase transition 
between the critical and the localized phase in the two-bath spin-boson model for $h_\ell = 0$, $s=0.8$, and as a function of the spin-boson coupling $\alpha$. For these parameters, the critical coupling
%\comment{\st{$\alpha_\mathrm{c}^\mathrm{MPS} = 0.765058$}} 
has been determined
using a variational MPS approach that is based on a Wilson chain with a logarithmic discretization for the bosonic bath
\cite{PhysRevLett.108.160401,PhysRevB.90.245130}. We show that our QMC method reaches
low-enough temperatures to distinguish the two phases 
via the spin susceptibility and get a precise estimate of the critical coupling from a finite-size analysis.

\begin{figure}
  \includegraphics[width=\linewidth]{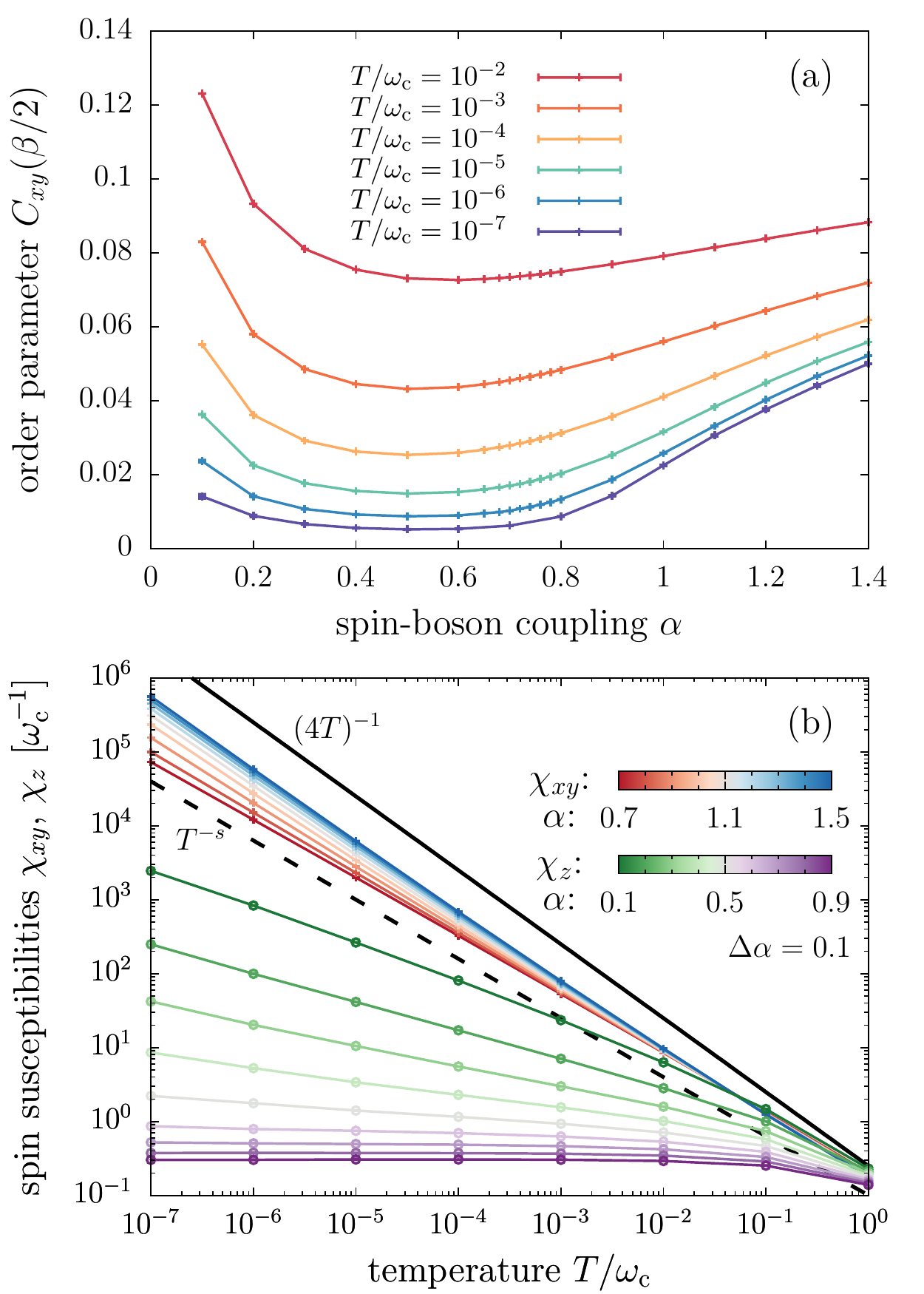}
  \caption{\label{fig:results_sus}%
(a)~Order parameter $C_{xy}(\beta/2)$ as a function of the spin-boson coupling $\alpha$ for different temperatures. In the localized phase,
$\lim_{\beta \to \infty} C_{xy}(\beta/2) = m^2_{\mathrm{loc}}$
converges to a finite local moment, whereas $m_\mathrm{loc}=0$
in the critical phase.
(b)~Spin susceptibilities $\chi_{xy}$ and $\chi_z$ as a function of temperature for different $\alpha$.
We show $\chi_{xy}$ ($\chi_z$) for $\alpha \in [0.7,1.5]$ ($\alpha \in [0.1,0.9]$) with
step size $\Delta\alpha =0.1$ as indicated by the color bars. The black line
corresponds to a free local moment with $\chi = (4T)^{-1}$
and the dashed line indicates the asymptotic behavior in the
critical phase, $\chi \sim T^{-s}$.
Here, $s=0.8$. 
  }
\end{figure}

For bath exponents $0<s<1$, the perturbative renormalization group has predicted an intermediate-coupling
fixed point where partial screening of the impurity leads to a critical phase
with power-law spin-spin correlations
$C_{xy}(\tau) \sim 1/\tau^{1-s}$
\cite{Sachdev2479, PhysRevB.61.4041, PhysRevB.66.024426, PhysRevB.66.024427, PhysRevB.61.15152}.
Numerical simulations revealed that the critical phase is only stable for $s^* <s <1$ [$s^* \approx 0.76$]
and $\alpha < \alpha_\mathrm{c}$ beyond which a local moment is formed that fulfills
$\lim_{\tau \to \infty} C_{xy}(\tau) = m_\mathrm{loc}^2>0$ \cite{PhysRevLett.108.160401,PhysRevB.90.245130}.
For a detailed discussion of the phase diagram, the fixed-point structure, and the critical properties
of the two-bath spin-boson model see Ref.~\onlinecite{PhysRevB.90.245130}. The QMC method developed in this paper allows us to approach the two
phases from finite temperatures. 
Figure~\ref{fig:results_sus}(a) illustrates 
the emergence of a local moment in the order parameter $C_{xy}(\beta/2)$
for $\alpha > \alpha_\mathrm{c}$, whereas $C_{xy}(\beta/2)$ scales to zero for $\alpha < \alpha_\mathrm{c}$.
The critical and the localized phases can also be distinguished by the low-temperature response
of the static spin susceptibilities $\chi_{xy}$ and $\chi_z$, as shown in Fig.~\ref{fig:results_sus}(b).
For $\alpha=0$ or $T/\omega_\mathrm{c} \gg 1$, we have $\chi_{xy} = \chi_{z} = 1/4T$.
For low-enough temperatures and deep in the localized phase,
$\chi_{xy}$ approaches a Curie law, $\chi_{xy} = m_\mathrm{loc}^2/T$,
with a finite local moment $m_\mathrm{loc}$. 
When entering the critical phase, $\chi_{xy}$ clearly deviates from the Curie behavior
and slowly converges to $\chi_{xy} \sim T^{-s}$; we do not show $\chi_{xy}$ for $\alpha<0.7$ because
all graphs are parallel to $\chi_{xy}$ at $\alpha=0.7$ but intersect with the others which reduces their
distinguishability.
On the other side, $\chi_z$ also shows a power-law dependence
with an exponent that is steadily reduced as $\alpha$ increases and eventually becomes zero
in the localized phase. Although our QMC method reaches temperatures as low as $T/\omega_\mathrm{c} = 10^{-7}$,
$C_{xy}(\beta/2)$ and $\chi_{xy}$ still experience finite-temperature effects which leads to
a slow convergence towards the expected $T\to 0$ behavior.

\begin{figure}
  \includegraphics[width=\linewidth]{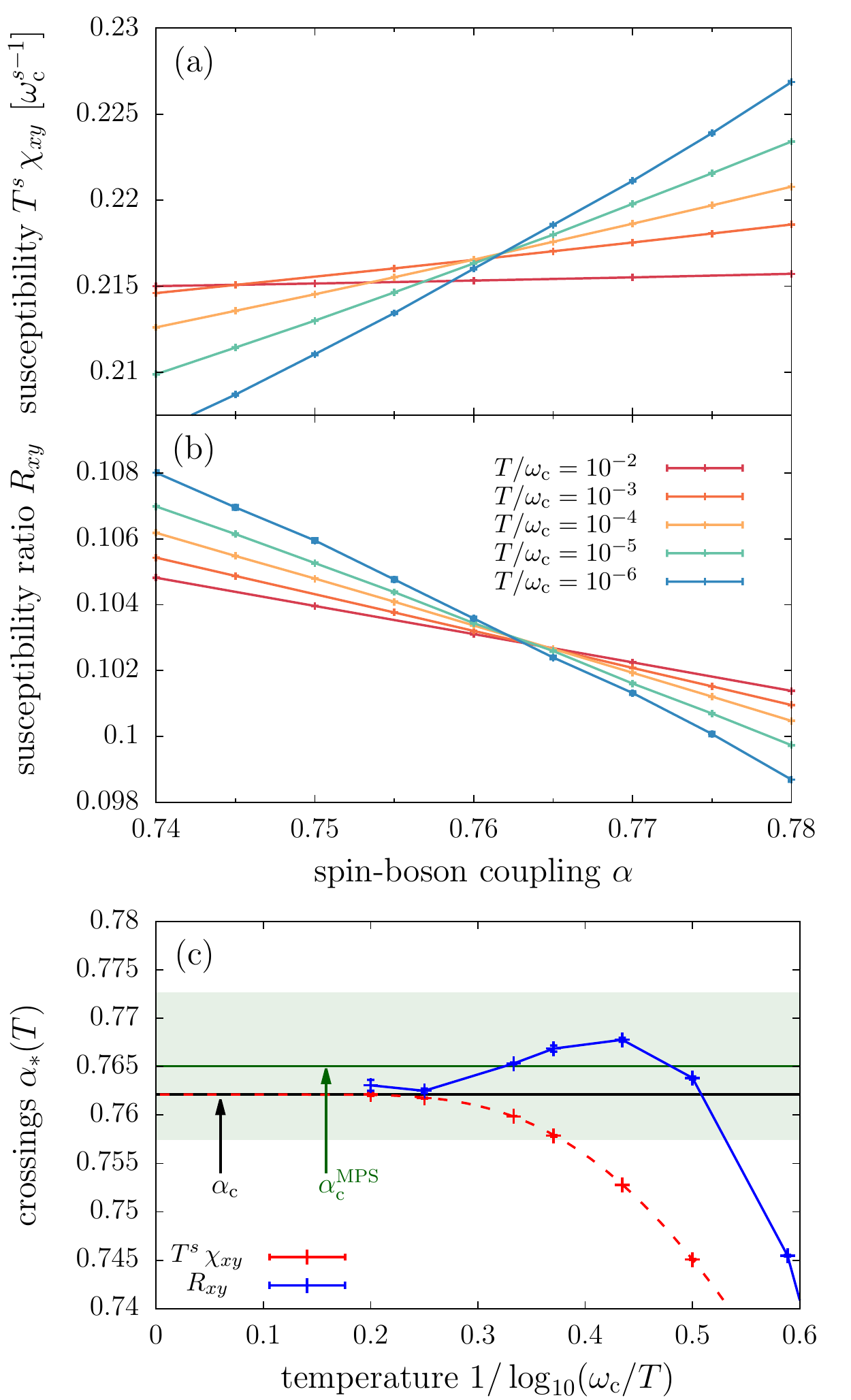}
  \caption{\label{fig:results_crossing}%
Finite-temperature analysis of (a) the rescaled susceptibility $T^s \chi_{xy}$
and (b) the susceptibility ratio $R_{xy}$ as a function of the spin-boson coupling $\alpha$.
(c) Temperature dependence of the crossings $\alpha_*(T)$ between pairs of data sets in (a) and (b)
with temperatures $\{T,T/10\}$.
We estimate the critical coupling as
$\alpha_\mathrm{c} = 0.76213(6)$
using a power-law fit of the form $\alpha_*(T) = \alpha_\mathrm{c} + A \, T^e$ for $T^s \chi_{xy}$ (red dashed line).
It is in good agreement with the variational MPS result
of Ref.~\onlinecite{PhysRevB.90.245130} which has a systematic error of up to $1\%$, indicated by the shaded area.
Here, $s=0.8$.
  }
\end{figure}
Our QMC method is powerful enough to determine a precise
estimate of the critical coupling. To this end, Figs.~\ref{fig:results_crossing}(a) and \ref{fig:results_crossing}(b) show the
rescaled susceptibility $T^s \chi_{xy}$ and the susceptibility ratio
\begin{align}
R_{xy}
	=
	\frac{\chi_{xy}(\im \Omega_1) }{ \chi_{xy}(\im \Omega_0)} \, ,
\end{align}
respectively. The latter is a generalization of the correlation ratio and its inverse can be related
to a finite-size estimator of the correlation length in imaginary time \cite{PhysRevB.100.014439}.
For $T\to 0$, both observables remain finite within the critical phase, whereas
$T^s \chi_{xy}$ diverges and $R_{xy}$ goes to zero in the localized phase. We can estimate
the critical coupling from a finite-size analysis of $T^s \chi_{xy}$ and $R_{xy}$;
here the inverse temperature plays the role of the system size, but in the imaginary-time direction.
For each pair of data sets with temperatures $\{T,T/10\}$, we extract the crossing points $\alpha_*(T)$
and collect them in Fig.~\ref{fig:results_crossing}(c). 
For low-enough temperatures, the crossings are expected to follow a power-law behavior $\alpha_*(T) = \alpha_\mathrm{c} + A \, T^e$,
from which we can estimate $\alpha_\mathrm{c}$ using a least-squares fit. From the crossings of $T^s \chi_{xy}$
we estimate $\alpha_\mathrm{c} = 0.76213(6)$; the nonmonotonic
temperature dependence of $\alpha_*(T)$ for $R_{xy}$ does not allow for a reliable fit, but the data is consistent with the estimate
from $T^s \chi_{xy}$.
%From the low-temperature behavior of $\alpha_*(T)$
%for both observables,
%we estimate $\alpha_\mathrm{c} =0.7625(10)$.
%Note that
%$R_{xy}$ converges faster to $\alpha_\mathrm{c}$ than $T^{s} \chi_{xy}$ but shows a nonmonotonic temperature
%dependence. 
Consistent crossings can also be obtained from 
the order-parameter ratio $C_{xy}(\beta/2) / C_{xy}(\beta/4)$, but larger statistical errors prohibited a precise estimation
of $\alpha_\mathrm{c}$. Our estimate of the critical coupling is in good agreement with the previous
MPS result $\alpha_\mathrm{c}^\mathrm{MPS} = 0.765058$ \cite{PhysRevB.90.245130};
%small deviations of 
note that the latter has a systematic error of
up to $1\%$ due to the logarithmic discretization of the bosonic bath in the Wilson chain \cite{PhysRevB.90.245130}.
Taking this systematic MPS error into account, our QMC estimate gives a more precise estimate of the critical coupling.
In a follow-up study \cite{2022arXiv220302518W}, we also determine the correlation-length exponent at this
quantum phase transition, which is in excellent agreement with the MPS result.

Finally, we want to emphasize again that the wormhole updates and the efficient sampling of the boson propagator
allow us to reach temperatures as low as $T/\omega_\mathrm{c} = 10^{-7}$---this is several orders of
magnitude lower than in previous QMC studies which only reached $T/\omega_\mathrm{c} \approx 10^{-4}$
\cite{PhysRevB.87.125102,PhysRevB.100.014439}. Based on our algorithmic developments, we are able
to resolve the characteristic low-temperature behavior of the critical and the localized phases in
Fig.~\ref{fig:results_sus}, in particular the difference between $\chi_{xy} \sim T^{-s}$ and $\chi_{xy} \sim T^{-1}$,
and get a precise estimate of the critical coupling in Fig.~\ref{fig:results_crossing}.
A systematic study of the closely-related SU(2)-symmetric spin-boson will be presented in Ref.~\onlinecite{2022arXiv220302518W}.
The directed-loop
updates work efficiently for all parameter regimes considered in this paper, because the average length of the loops
is given by $\chi_{xy}$ and therefore diverges with the characteristic temperature scale of the problem.
For the lowest temperatures and strongest couplings, we have reached expansion orders of approximately
20 million vertices.

\section{Generalizations to other interaction types\label{Sec:FurtherInteractions}}

We have introduced our QMC method for the XXZ spin-boson model
and the Jaynes-Cummings model. In the following, we discuss
how other interaction types can be included for impurity models
and how the wormhole updates can be applied to lattice models.

\subsection{Impurity models}

\subsubsection{Magnetic field in the $xy$ plane}

For the impurity models discussed before, it might be of interest to apply a magnetic field
also in the $xy$ plane. We can decompose the additional field as follows:
\begin{align}
\hat{H}_{h_{xy}}
&=
-h_x \spinx - h_y \spiny \nonumber \\
&=
- \frac{1}{2} \left[ \left( h_x - \im h_y \right) \spinp
+ \left( h_x + \im h_y \right) \spinm \right] \, .
\end{align}
Because the retarded spin-boson interactions in Eqs.~(\ref{Eq:SretXXZ}) and (\ref{Eq:SretJC})
always contain a pair of 
$\spinp$ and $\spinm$ operators, the magnetic-field terms
in the perturbation expansion have to appear in pairs as well and therefore
do not lead to a sign problem.

We can treat the $xy$ magnetic field as an additional interaction type.
To formulate a full updating procedure, we add the diagonal term
$- C  \int d\tau \, \hat{\mathbb{1}}(\tau)$
to the interaction vertex. It is convenient to choose the constant
prefactor equal to $h_{xy}/2$ where $h_{xy} = \sqrt{h_x^2 + h_y^2}$.  Then,
all vertex weights have the same absolute value, which simplifies the
solution of the directed-loop equations. If the directed loop enters
one of these vertices, we want to transform a unit operator into a spin-flip operator
or vice versa which prohibits the loop head from propagating any further locally. However,
we can choose a random magnetic-field vertex in our world-line configuration (including the original vertex)
and continue the construction of the loop from there.
The exit leg has to be chosen in such a way that it is consistent with the propagating operator
and the world-line configuration. The construction of the loop is completed when the loop head returns
to its original starting point. Including the magnetic field in this way has the advantage
that we can measure $\expvtext{h_x \spinx + h_y \spiny} = \expvtext{n_{xy,1}}/\beta$
from the number of off-diagonal magnetic-field vertices in the perturbation expansion.

\subsubsection{XYZ spin-boson model}

Our discussion of the spin-boson model in Eq.~(\ref{Eq:HamXYZ}) was restricted
to the XXZ case. Here, we want to mention that one can also simulate the full problem
with three independent couplings $\lambda_\ell$.
Then, the off-diagonal vertex in Eq.~(\ref{Eq:h1XXZ}) has to be replaced by
\begin{align}
\nonumber
\hat{h}_1(\tau,\tau')
	=
	 & \frac{\lambda_{x}+\lambda_{y}}{4} \left[ \spinp(\tau) \, \spinm(\tau') + \spinm(\tau) \, \spinp(\tau') \right] \\ %\, ,
	 + & \frac{\lambda_{x}-\lambda_{y}}{4} \left[ \spinp(\tau) \, \spinp(\tau') + \spinm(\tau) \, \spinm(\tau') \right]  \, .
	\label{Eq:h1XYZ}
\end{align}
The new terms $\spinp(\tau) \, \spinp(\tau')$ and $\spinm(\tau) \, \spinm(\tau')$ have to be considered
as additional vertex types $v$, \ie, $v=7$ and $v=8$. With these vertex types present, the directed loop can exit
all four legs of a vertex and the solution of the directed-loop equations  becomes a bit more complicated
than before. As long as we do not apply an additional field in the $xy$ plane, the second term in Eq.~(\ref{Eq:h1XYZ})
will not lead to negative weights because it always has to appear in pairs in the perturbation expansion.

\subsubsection{Original spin-boson model}

For completeness, we also want to mention how to simulate
the original spin-boson model in Eq.~(\ref{Eq:Ham_spin-boson}).
Because the bosonic bath only couples to the $z$ component of the spin,
we cannot immediately apply the wormhole updates as defined above.
To obtain a retarded spin-flip interaction, we can formulate our QMC
method in the $\spinx$ basis. This corresponds to a rotation of the
spin operators using the Hadamard matrix (which exchanges $\spinx \leftrightarrow \spinz$
in the absence of $\spiny$).
The interaction part becomes
\begin{gather}
\nonumber
\hat{h}_1(\tau,\tau')
	=
	 \frac{\lambda_{z}}{4} \, \big[ \spinp(\tau) \,  \spinm(\tau') + \spinm(\tau) \, \spinp(\tau') 
	 \\
	  \qquad \qquad\qquad\qquad + \spinp(\tau) \,  \spinp(\tau') + \spinm(\tau) \, \spinm(\tau')
	 \big] \, ,
\\
\hat{h}_2(\tau,\tau')
	=
	C +  \frac{h_x}{2} \left[\spinz(\tau) + \spinz(\tau')\right] \, ,
\end{gather}
and leads to positive Monte Carlo weights for $C\geq \absolute{h_x}/2$.
The off-diagonal term now contains two additional vertex types.
As a result, the directed loops can exit all of the four vertex legs
and the directed-loop equations can be solved using
linear-programming techniques \cite{PhysRevE.71.036706}.

For the spin-boson model, a cluster algorithm had already been
formulated in Ref.~\onlinecite{PhysRevLett.102.030601} in terms
of a retarded interaction between $\spinz$ operators. The spin-boson
model has close similarities with the long-range Ising model in
a transverse magnetic field which can be simulated efficiently
in the SSE representation \cite{PhysRevE.68.056701}. It
might also be possible to extend this method to long-range interactions
in imaginary time.

\subsection{Lattice models}

We have developed our QMC method for a single spin coupled
to a dissipative bosonic bath, but the wormhole updates trivially
extend to lattice models. In the simplest case, each lattice site
is coupled to an independent bath, so that the interaction vertex
only gets an additional site index $i$. The details of our method
stay exactly the same, we only have to draw a random $i \in [1, L]$
when proposing the addition of a diagonal vertex, which leads to an
extra factor of $1/L$ in Eq.~(\ref{Eq:DiagProposeAdd}).
To couple the spins at different lattice sites, we can include any
interaction vertex that can be simulated
in the SSE representation without a sign problem, \eg,
an XXZ spin-exchange coupling between nearest neighbors.
The additional vertex $\ham{}$ transfers into the interaction picture
as $\S = \int d\tau \, \ham{}(\tau)$.
The spin operators obtain dummy time variables
that are sampled during the diagonal updates to determine their position
in the world-line configuration.
The diagonal
updates have to be formulated using the Metropolis scheme in Sec.~\ref{Sec:diagonalupdates},
whereas the probability tables for the directed-loop updates stay the same.
Note that the vertex structure need not be the same for different interaction types,
as long as the loop updates can be formulated for each type individually.
Eventually, the wormhole updates introduced in this paper allow us to study the effects of
dissipation on a variety of lattice models. While the worm algorithm can
be applied when a diagonal operator couples to a bosonic bath \cite{PhysRevLett.113.260403},
our formulation with directed loops allows us to simulate more generic couplings that
also include off-diagonal terms. Results for a Heisenberg chain coupled 
to an ohmic bath will be presented elsewhere \cite{ohmicbath}.

We have introduced the wormhole updates for boson-mediated interactions that are local in space
but nonlocal in imaginary time. In general, the wormhole updates can be applied
to long-range interactions in space or time, as long as the Monte Carlo weights
are positive. Examples arise in quantum optics where global bosonic modes couple to the spins
at different sites, \eg, in the Dicke model \cite{https://doi.org/10.1002/qute.201800043}
or in trapped-ion simulators \cite{PhysRevLett.92.207901}---for the latter, the absence
of a sign problem has to be assessed from case to case. 
However, we expect that polariton models like the
Jaynes-Cummings-Hubbard model will introduce a sign problem in our formulation
because integrating out the bosonic hoppings $\bcr{i} \ban{j}$ leads to a nonlocal
boson propagator that has negative contributions for $i \neq j$ \cite{PhysRevB.91.245147}; 
in that case,
it is better to simulate the bosons directly \cite{PhysRevA.80.033612,PhysRevA.84.041608}.
Moreover, we can design long-range interactions in space and time that
are not related to a spin-boson Hamiltonian but can be simulated
with the wormhole updates.

\section{Conclusions\label{Sec:Conclusions}}

We developed an exact QMC method to simulate spin systems coupled to dissipative bosonic baths in terms of retarded spin interactions.
Our method makes use of the directed-loop updates which were originally formulated in the SSE representation \cite{PhysRevE.66.046701}
and recently generalized to retarded interactions \cite{PhysRevLett.119.097401}. To include the retarded spin-flip interactions into the
framework of the directed-loop updates, we introduced the nonlocal wormhole updates which allow the loop to tunnel between the two operators
of a vertex. The formulation of the directed-loop updates remains as simple as in the SSE representation \cite{PhysRevE.66.046701}
since the time-dependence of the retarded interaction is fully accounted for in the diagonal updates \cite{PhysRevLett.119.097401}.
Our method applies to any spectral distribution of the bath, which can be sampled efficiently---along with the time-dependence of the boson propagator---during the diagonal updates.
The ideas developed in this paper are also applicable to related QMC methods like the worm algorithm \cite{Prokofev:1998aa}
which make use of worm/directed-loop updates to sample world-line configurations.

We demonstrated the efficiency of our QMC method for the two-bath spin-boson model.
Our method reached significantly lower temperatures than previous QMC approaches \cite{PhysRevB.87.125102,PhysRevB.100.014439},
which allowed us to identify the characteristic finite-temperature response of the spin susceptibility in the critical and the localized phase.
To estimate the quantum critical coupling between these two phases,
we compared the finite-size behavior of different observables.
Our estimated $\alpha_\mathrm{c}$ is in good agreement with previous results from a variational MPS study
\cite{PhysRevLett.108.160401,PhysRevB.90.245130}.

In future studies, our QMC method will enable us to determine the phase diagrams
of spin-boson models coupled to multiple baths \cite{2022arXiv220302518W}. Open questions include the quantum
critical properties of these models and their relation to spin systems with long-range interactions in space.
Spin-boson interactions also play an important role in the more complex Bose-Fermi Kondo model which
is a central model for our understanding of the quantum critical behavior appearing in heavy-fermion
systems \cite{Si:2001aa}.
For the local spin-boson couplings considered in this paper,
the wormhole updates can be combined 
with any lattice model that can be simulated already in the SSE representation, which
will allow us to study the effects of dissipation on finite systems \cite{ohmicbath}.
We expect that our QMC method will be able to simulate more general lattice models
coupled to bosonic modes which describe light-matter interactions in quantum optics
or trapped-ion systems---spin-boson models can be realized, \eg, in quantum simulators 
\cite{PhysRevA.78.010101}.
In general, the wormhole updates not only apply to long-range interactions in imaginary time but also in space.
Our algorithmic developments therefore open up a route to study a completely new class of models within the framework of the well-established
directed-loop algorithm.

\acknowledgments

I thank F.~Assaad, M.~Hohenadler, D.~Luitz,
F.~Parisen Toldin, and M.~Vojta for helpful discussions.

\appendix*

\section{Implementation\label{Sec:Implementation}}

The implementation of the directed-loop algorithm in the SSE representation
is well documented and we refer to Refs.~\onlinecite{PhysRevE.66.046701, doi:10.1063/1.3518900}
for a detailed description of the necessary data structures.
In the following, we will give a quick overview over the modifications that become necessary
in the interaction representation due to the nonlocality in imaginary time.

In the SSE representation, the vertices are saved in an operator string with a fixed length
that is traversed sequentially during the diagonal updates; in this process, unit operators are exchanged
with diagonal ones and vice versa. In our formulation, the vertices are saved in an unsorted list where each
element has a structure that contains all vertex variables. For the spin-boson vertices defined in this
paper, each vertex has variables $\nu = \{\typeint, v, \omega, \tau, \tau' \}$; note that it is more useful
to save the vertex type $v$ than the operator type $a$.
Before we start a block of diagonal updates,
we once go through the vertex list and set up a new list that includes all time variables
at lattice site $i$ where an off-diagonal operator
is applied. After sorting this list, we can access the spin configuration at each position in the world-line configuration
from the knowledge of the initial state $\ket{\alpha}$. Although sorting algorithms require $\mathcal{O}(\beta \log\beta)$
operations for each site and are mathematically more expensive than the directed-loop updates, the construction of the loops
remained the most expensive task for all temperatures accessible to our simulations. To add a diagonal vertex, we can
access the spin configurations from the sorted list, draw the vertex variables as described in 
Sec.~\ref{Sec:diagonalupdates}, calculate the Metropolis acceptance, and eventually add the vertex as element $n+1$ of the vertex list. To remove a diagonal vertex, we randomly pick a diagonal vertex, exchange it with element $n$ saved in the vertex list, and then reduce the integer $n$ that counts the number of vertices by one. Although the diagonal updates change the number
of vertices, we can work with a fixed length for the vertex list and only take into account the first $n$ elements.

The implementation of the directed-loop updates is very similar to Ref.~\onlinecite{PhysRevE.66.046701}.
To create the doubly-linked vertex list, we first traverse through the vertex list again and make separate
lists for the time variables and subvertex labels---the latter is a combination of the vertex number
and a label for operator 1 or 2. We create an index list for the time variables to sort the list of subvertex labels.
From this we can setup the linked vertex list. The construction of the directed loops then follows the description
in Ref.~\onlinecite{PhysRevE.66.046701}. When we enter a vertex of type $v$ at leg $l_i$, we can determine
the exit leg $l_e$ from a probability table. Even for the wormhole moves, the combination of the vertex variable
and the exit leg exactly determines the label of the exit vertex in the linked vertex list. During the construction of the loop,
we also update the vertex configurations in the vertex list. Note that the correct measurement of the off-diagonal
time-displaced correlation functions requires that we choose a random starting time for the directed loop. After
we have finished a block of directed-loop updates, we also update the initial state $\ket{\alpha}$, as described
in Ref.~\onlinecite{PhysRevE.66.046701}.

The number of proposals during the diagonal updates and the number of constructed loops in the directed-loop updates
get adjusted during the Monte Carlo warmup---in such a way that every vertex is touched at least once on average---and remain
fixed afterwards. We start the warmup from a randomly chosen $\ket{\alpha}$ and an empty vertex list. To speed up
the warmup procedure, we use a beta-doubling scheme. For the measurement of certain observables, we might
have to sort parts of the vertices again.

\end{document}